\documentclass[lettersize,journal]{IEEEtran}
\usepackage{amsmath,amsfonts}
\usepackage{array}
\usepackage[caption=false,font=normalsize,labelfont=sf,textfont=sf]{subfig}
\usepackage{textcomp}
\usepackage{stfloats}
\usepackage{url}
\usepackage{verbatim}
\usepackage{graphicx}

\usepackage{xspace}

\usepackage{algorithmicx}
\usepackage{algpseudocode}

\usepackage{multirow} 
\usepackage{array}    
\usepackage{booktabs} 
\usepackage{makecell} 

\def\BibTeX{{\rm B\kern-.05em{\sc i\kern-.025em b}\kern-.08em
    T\kern-.1667em\lower.7ex\hbox{E}\kern-.125emX}}
\usepackage{balance}
\begin{document}
\author{IEEE Publication Technology Department


\author{\IEEEauthorblockN{
Jae-Young Kim
\IEEEmembership{Graduate Student Member, IEEE}, 
Donghyuk Kim
\IEEEmembership{Graduate Student Member, IEEE},
Seungjae Yoo
\IEEEmembership{Graduate Student Member, IEEE},
Sungyeob Yoo
\IEEEmembership{Graduate Student Member, IEEE},
Teokkyu Suh
\IEEEmembership{Graduate Student Member, IEEE},
Joo-Young Kim
\IEEEmembership{Senior Member, IEEE}}\\
\IEEEauthorblockA{School of Electrical Engineering,
Korea Advanced Institute of Science and Technology (KAIST),
Daejeon, Republic of Korea}\\
\{jykim1109, kar02040, goldenyoo, sungyeob.yoo, ejrrb102, jooyoung1203\}@kaist.ac.kr
}

\thanks{Manuscript created October, 2020; This work was developed by the IEEE Publication Technology Department. This work is distributed under the \LaTeX \ Project Public License (LPPL) ( http://www.latex-project.org/ ) version 1.3. A copy of the LPPL, version 1.3, is included in the base \LaTeX \ documentation of all distributions of \LaTeX \ released 2003/12/01 or later. The opinions expressed here are entirely that of the author. No warranty is expressed or implied. User assumes all risk.}}


\newcommand{\sysname}{RED\xspace}
\title{\sysname: Energy Optimization Framework for eDRAM-based PIM with Reconfigurable Voltage Swing and Retention-aware Scheduling} 

\maketitle

\begin{abstract}
In the era of artificial intelligence (AI), Transformer demonstrates its performance across various applications. The excessive amount of parameters incurs high latency and energy overhead when processed in the von Neumann architecture. Processing-in-memory (PIM) has shown the potential in accelerating data-intensive applications by reducing data movement. While previous works mainly optimize the computational part of PIM to enhance energy efficiency, the importance of memory design, which consumes the most power in PIM, has been rather neglected.

In this work, we present RED, an energy optimization framework for eDRAM-based PIM. We first analyze the PIM operations in eDRAM, obtaining two key observations: 1) memory access energy consumption is predominant in PIM, and 2) read bitline (RBL) voltage swing, sense amplifier power, and retention time are in trade-off relations. Leveraging them, we propose a novel reconfigurable eDRAM and retention-aware scheduling that minimizes the runtime energy consumption of the eDRAM macro. The framework pinpoints the optimal operating point by pre-estimating energy consumption across all possible tiling schemes and memory operations. Then, the reconfigurable eDRAM controls the RBL voltage swing at runtime according to the scheduling, optimizing the memory access power. Moreover, RED employs refresh skipping and sense amplifier power gating to mitigate the energy consumption overhead coming from the trade-off relation. Finally, the RED framework achieves up to 3.05$\times$ higher energy efficiency than the prior SRAM-based PIM, reducing the energy consumption of eDRAM macro up to 74.88\% with reconfigurable eDRAM and optimization schemes, requiring only 3.5\% area and 0.77\% energy overhead for scheduling.
\end{abstract}


\section{Introduction}
\label{introduction}

In the era of artificial intelligence (AI), Transformer has made a significant impact over convolutional neural networks (CNN) by demonstrating its performance across a variety of applications, including generative models ~\cite{radford2019language}, natural language processing ~\cite{devlin2018bert}, and image processing ~\cite{liu2022video}. Although transformer-based models provide remarkable results, many studies show that the excessive amount of parameters causes inefficiency when computed in the traditional von Neumann computer architecture. Therefore, many processing-in-memory (PIM) based accelerators have been proposed to accelerate Transformer: \cite{zhou2022transpim, tu2022trancim,kim2022overview, PIM_TRANS_01, PIM_TRANS_02, PIM_TRANS_03, PIM_TRANS_04}.

By placing the processing unit in/near the memory, the high latency and power consumption caused by data movement can be significantly reduced, making PIM superior for accelerating data-intensive applications. 
By leveraging the benefits of PIM, various fully customized designs based on SRAM have been proposed with its high reliability~\cite{kim2020z, heo2022t, kim2023sp, SRAM_PIM_01, SRAM_PIM_02, SRAM_PIM_03, SRAM_PIM_04, SRAM_PIM_05, SRAM_PIM_06}. However, due to the increasing demand for higher memory capacity with increasing model sizes, PIM designs using the cell with higher density, such as embedded DRAM (eDRAM), have been recently proposed \cite{kim2023dynaplasia, xie2022gain, eDRAM_PIM_01, eDRAM_PIM_02, eDRAM_PIM_03} to provide better area and power efficiency compared to SRAM-based approach.

PIM achieves high throughput by executing multiple PIM macros in parallel. As PIM macros carry out most of the operations, they account for most of the power consumption, about 70\%, as shown in the power breakdown of \cite{tu2022trancim}. Consequently, designing an energy-efficient PIM macro is vital for boosting overall energy efficiency. To accomplish this, previous works mainly focus on optimizing the computational part. To mitigate the area and power overhead that a processing unit imposes on the memory, they have introduced various techniques in computation such as zero-skipping \cite{kim2020z}, approximate computing \cite{wang2022dimc}, and optimized adder tree \cite{chih202116}. However, the importance of memory design, which consumes the most power in PIM operation, has been rather neglected.

\begin{figure}[t]
\centering
\includegraphics[width=0.48\textwidth]{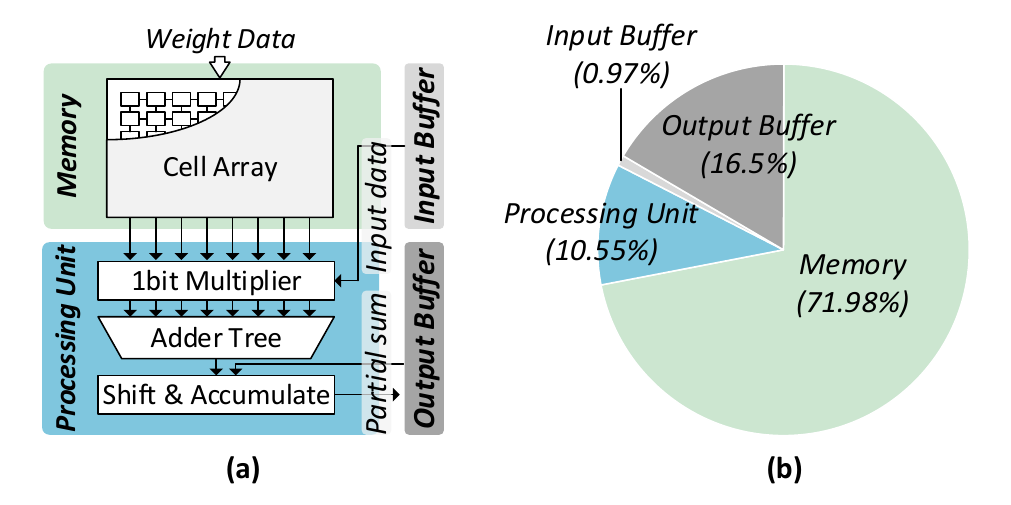}
\caption{(a) PIM Macro (b) Power Consumption Breakdown}
\vspace{-0.1in}
\label{figure1}
\end{figure}

Figure~\ref{figure1} (a) presents a PIM macro designed to accelerate general matrix-matrix multiplication (GEMM) or general matrix-vector multiplication (GEMV). This macro includes an input buffer, an output buffer, a memory, a 1-bit multiplier, an adder tree, and an accumulator. The multiplier and adder tree facilitate the multiply-accumulate (MAC) operations, and the accumulator merges the resulting partial sum. Figure~\ref{figure1} (b) shows the power consumption breakdown for each unit. As the breakdown shows, memory access consumes 71.98\% of the total power during an operation. This finding underscores the importance of optimizing memory access power to develop an energy-efficient PIM macro.

In this work, we propose \sysname, an energy optimization framework for eDRAM-based PIM. 
First, we make two pivotal observations to attain high energy efficiency in eDRAM-based PIM: 1) energy consumption of memory access is dominant in PIM; and 2) a trade-off exists among read bitline (RBL) voltage swing, sense amplifier power, and retention time. We find that optimizing memory access power without considering a specific use case is an inefficient approach. Leveraging these insights, we propose a novel reconfigurable eDRAM and retention-aware scheduling. Through the \sysname's retention-aware scheduling, our framework identifies the optimal tiling scheme and memory operation. Then, by controlling the memory access pattern and memory operation based on the outcomes of the scheduling, the power and energy consumption of memory in PIM operation is fully optimized. 

With these approaches, the \sysname framework achieves up to 2.66$\times$, 3.05$\times$, and 8.16$\times$ higher energy efficiency than the worst case, Neural Cache~\cite{eckert2018neural}, and eDRAM baseline for various Transformer models. Our proposed retention-aware scheduler and controller require less than 3.5\% area overhead and 1\% energy overhead, promising effective solutions for the development of eDRAM-based PIM. We summarize the key contributions of our work below.
\begin{figure}[t]
\centering
\includegraphics[width=0.48\textwidth]{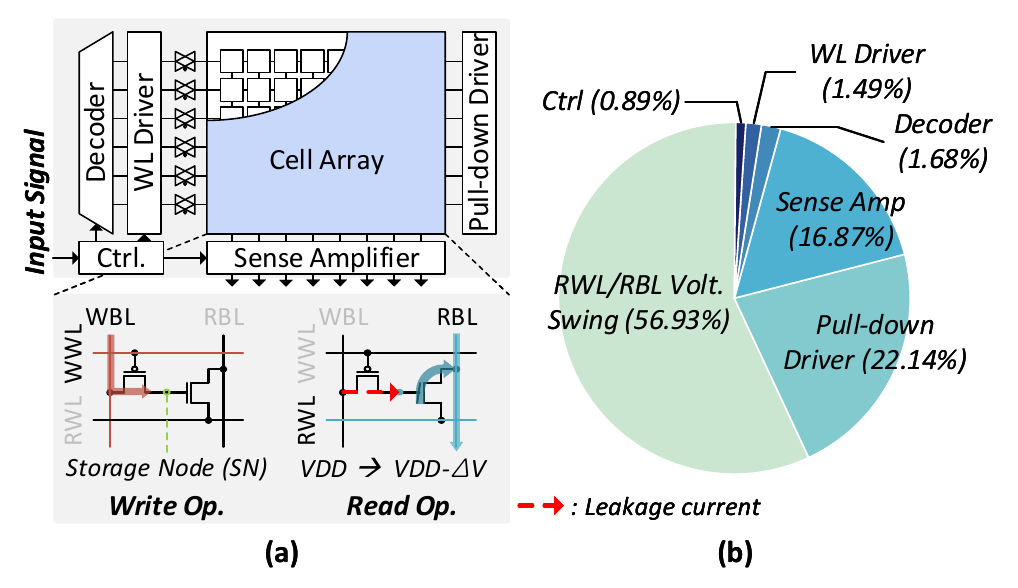}
\caption{(a) Memory and 2T eDRAM Cell Structure (b) Power Breakdown of eDRAM Access}
\label{figure2}
\end{figure}

\begin{itemize}
    \item We first analyze how the use case affects energy consumption of PIM and discuss the trade-off involved in the memory operation.
    \item We propose an energy optimization framework for eDRAM-based PIM. The \sysname framework fully optimizes the energy consumption of memory access, which is the main source of energy consumption in PIM architecture.
    \item We develop retention-aware scheduling that can identify the most energy-efficient tiling scheme and memory operation.
    \item We propose a novel reconfigurable eDRAM allowing optimal memory control tailored to the actual use case.
\end{itemize}

\section{Background}
\label{background}

Embedded DRAM (eDRAM) cell varies by the number of transistors and structure, such as 1T1C, 2T, 3T, and 4T. While 1T1C provides the highest memory density, due to its destructive read issue as described in~\cite{xie2022gain}, it is hard to utilize the multi-row activation scheme, which is commonly employed in PIM. Hence, this paper adopts a 2T eDRAM cell, which has high memory density and the capability for non-destructive read.


Figure~\ref{figure2} (a) depicts the structure of an eDRAM macro and a 2T eDRAM cell, along with the write and read operations. The macro consists of a controller, a decoder, a wordline (WL) driver, a cell array, a pull-down driver, and a sense amplifier. For write operation, data is stored in the storage node (SN) by driving write wordline (WWL) to a negative voltage (-$V_{th}$) and driving the data through write bitline (WBL). In a read operation, the pull-down driver drives read wordline (RWL) to ground (VSS). This operation causes read bitline (RBL), initially precharged to VDD, to fluctuate in response to a storage node (SN) voltage. In this work, the sensing margin denotes the voltage level difference between the reference voltage and data 0, 1. The sense amplifier then converts this RBL voltage swing to a digital value by comparing it with the reference voltage. However, as shown in  Figure~\ref{figure2} (a) with the red arrow, a leakage current flows through the PMOS transistor. This current changes the charge stored in the SN, leading the RBL voltage swing to deviate from the ideal case. The retention time means the duration until the sense amplifier accurately converts the data. Because of the non-ideality, eDRAM requires periodic data refresh, which imposes large energy and latency overhead at the system level. Figure~\ref{figure2} (b) shows the power breakdown of an eDRAM macro with 32$\times$512 cell array. Due to the high capacitance of long metal wires composing the cell array, the majority of power consumption is from RWL/RBL voltage swing, pull-down driver, and sense amplifier, occupying 56.93\%, 22.14\%, and 16.87\%, respectively. It implies that optimizing these components can maximally reduce the memory access power.
\begin{figure}[t]
\centering
\includegraphics[width=0.48\textwidth]{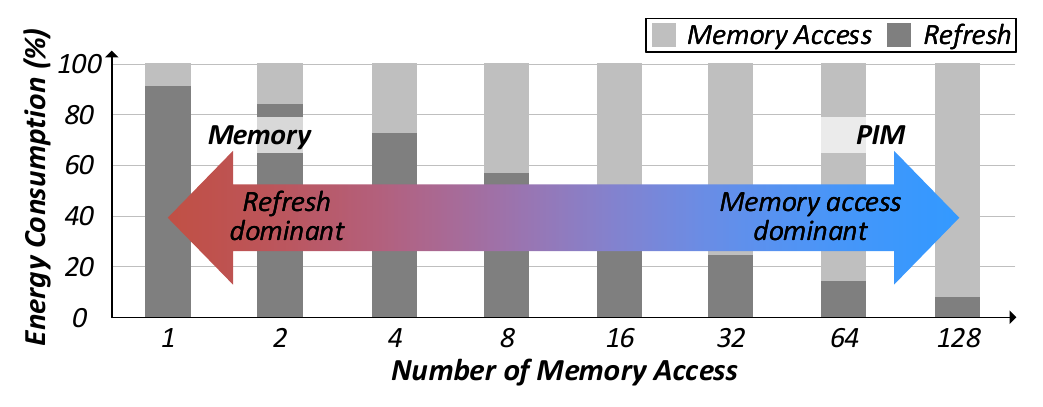}
\caption{Breakdown of Memory Energy Consumption on Different Use Cases}
\label{figure3-1}
\end{figure}

\section{Observations}
\label{observations}

In this work, we analyze the key observations outlined below. These insights form the foundation of our proposed \sysname framework to fully optimize both the power and energy consumption of memory access, achieving high energy efficiency.

\textbf{Observation 1: dominant energy consumption of memory access in PIM.} In the von Neumann architecture, off-chip data movement causes the primary energy consumption in accessing memory. To minimize this energy consumption, the architecture aims to maximally reuse data loaded in the processor with its on-chip memory. In contrast, PIM architecture reduces data movement overhead in exchange for much more frequent memory access for operations by in-memory processing units. The key distinction lies in memory utilization, making the difference in the circuit design of memory in each architecture. Unlike conventional architectures where the energy consumption of refresh is substantial, leading to designs focused on reducing this refresh overhead to enhance performance, PIM architecture faces a different challenge.

Figure ~\ref{figure3-1} shows the ratio of refresh and energy consumption of memory access according to the number of memory accesses where the lifetime of data is set as 1000$\mu$s. The memory is assumed as eDRAM with a retention time of around 100$\mu$s. The refresh overhead is dominant when the number of memory accesses is small, as in the conventional architecture use case. However, the energy consumption of memory access becomes dominant as the number of memory increases; that is the use case of PIM architecture. The graph emphasizes the necessity of designing memory considering a property of architecture and actual use case. Additionally, it is essential to optimize the energy consumption of memory access to improve performance in PIM effectively.


\begin{figure}[t]
\centering
\includegraphics[width=0.48\textwidth]{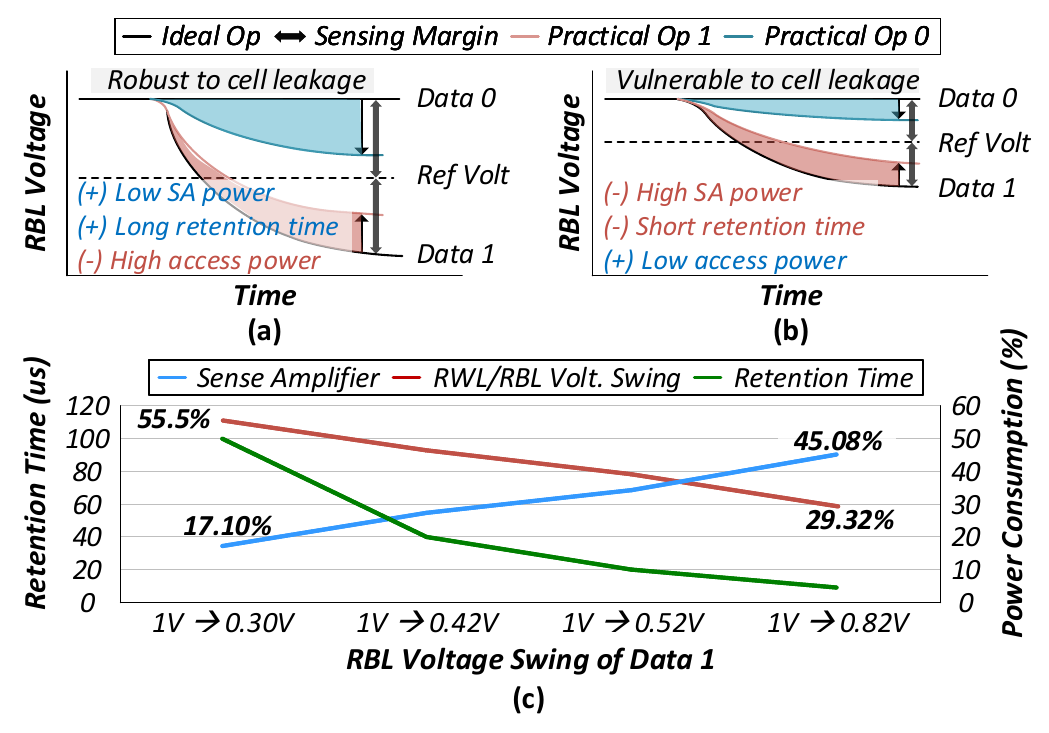}
\caption{(a) eDRAM Operation with Large RBL Voltage Swing (b) eDRAM Operation with Small RBL Voltage Swing (c) Trade-off Among RBL Voltage Swing, Sense Amplifier Power, and Retention Time Analysis}
\vspace{-0.1in}
\label{figure3-2}
\end{figure}

\textbf{Observation 2: trade-off among RBL voltage swing, sense amplifier, and retention time.} eDRAM operation with large RBL voltage swing, as shown in Figure ~\ref{figure3-2} (a), enhances the sensing margin. This enhancement allows the sense amplifier to convert the analog signal into a digital signal faster. As a result, the total current flowing through the sense amplifier diminishes, leading to the reduced power consumption of the sense amplifier  (17.10\%), as shown in Figure ~\ref{figure3-2} (c). Even though the charge in a SN fluctuates due to a leakage current, a large sensing margin ensures that the sense amplifier converts the data correctly, resulting in a longer retention time (100$\mu$s). However, as explained in Section ~\ref{background}, the memory access power increases because the RBL voltage swing is the main source of power consumption. Conversely, a smaller RBL voltage swing leads to higher power consumption of the sense amplifier (45.08\%), shortened retention time (9$\mu$s), and smaller memory access power, as depicted in Figure ~\ref{figure3-2} (b) and (c).

Simply opting for a small RBL voltage swing to optimize memory access power falls short of an optimal solution due to the inherent trade-off of eDRAM. While it does lower memory access power, it rather increases the power consumption of the sense amplifier and refresh overhead. Therefore, we have to select a memory operation considering the specific PIM use case that minimizes both memory access power and overall energy consumption.

\begin{figure}[t]
\centering
\includegraphics[width=0.48\textwidth]{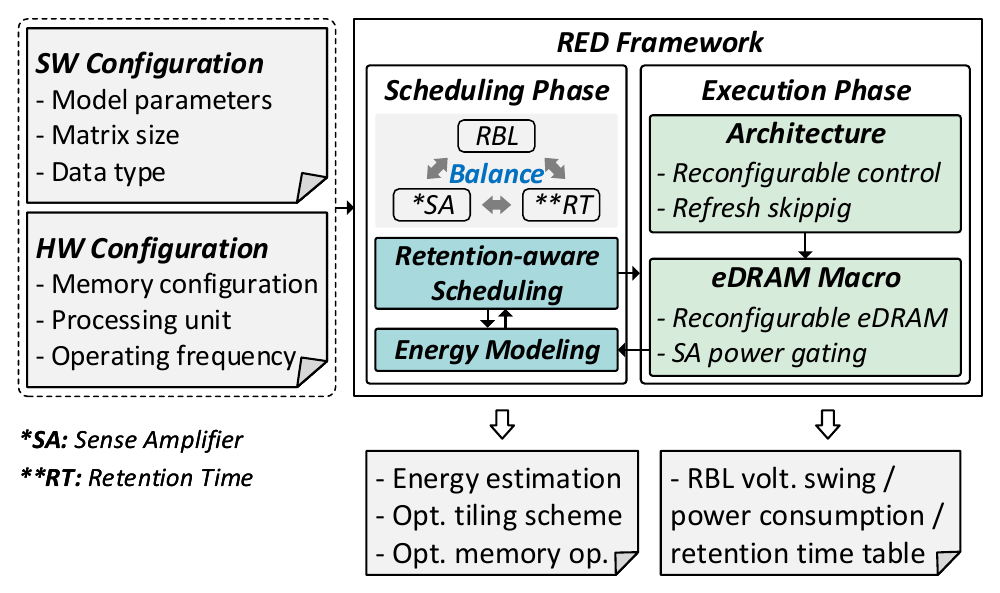}
\caption{Overview of The \sysname Framework}
\vspace{-0.1in}
\label{figure4}
\end{figure}

\section{\sysname Framework}
\label{framework}
We propose the \sysname framework, an energy optimization framework for eDRAM-based PIM that can fully optimize the energy consumption of the memory with retention-aware scheduling, optimal architecture, and reconfigurable circuit design.

Figure ~\ref{figure4} presents the overview of the \sysname framework. The framework receives a software configuration (e.g., model parameters, input and weight matrix shapes, data types, etc.) and a hardware configuration (e.g., memory configuration, processing units, operating frequency, etc.) as inputs and goes through two phases: \textit{the scheduling and execution phase}. These two phases collectively allow the \sysname framework to significantly reduce overall energy consumption by optimizing memory operations, which are the primary contributors to energy overhead in PIM operation. The \sysname framework can optimize energy consumption in PIM across diverse configurations. The following section delves into a detailed description of these two phases, revealing how they form the backbone of the \sysname framework.

\subsection{Retention-aware Scheduling}
\label{retention-aware scheduling}
As discussed in Section ~\ref{observations}, achieving high energy efficiency in PIM requires optimizing the energy consumption of memory access. However, simply reducing RBL voltage swing without tailoring it to a specific use case is not the most efficient approach. To address this problem, the \sysname’s scheduling phase undertakes two key tasks: 1) identifying the optimal tiling scheme and 2) selecting the optimal memory operation.

The eDRAM memory requires periodic refreshes due to its retention time. However, it is possible to skip refresh in two cases. Firstly, when the lifetime of data is shorter than the retention time, as proposed in \cite{tu2018rana}, there is no need for refresh. Secondly, refresh can be skipped for the data that is no longer used for computation. Given that PIM architecture cannot process the whole input and weight matrices simultaneously, it employs tiling for GEMM. Memory access pattern and output generation order vary with a tiling scheme, influencing the lifetime of mapped data. Therefore, to enhance the energy efficiency of eDRAM-based PIM, it is essential to schedule operations considering the lifetime and the retention time. 
Additionally, optimally balancing eDRAM's inherent trade-off by considering the actual use case can further reduce energy consumption.
The \sysname’s scheduling phase goes through the following four steps to determine the most optimal tiling scheme and memory operation for given input configurations.

\begin{table}[t]
\centering
\caption{\sysname's Variable Definition}
\includegraphics[width=0.48\textwidth]{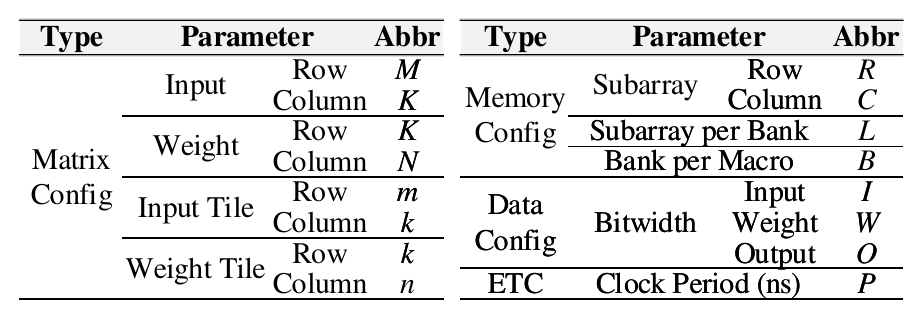}
\label{figure8}
\end{table}

\textbf{Step 1: Find all possible tiling schemes.} The scheduler identifies all possible tiling schemes (i.e., loop order and tile shape) based on the input configurations. Table 1 presents the variable definitions used by the \sysname's scheduling phase.

\begin{figure}[t]
\centering
\includegraphics[width=0.48\textwidth]{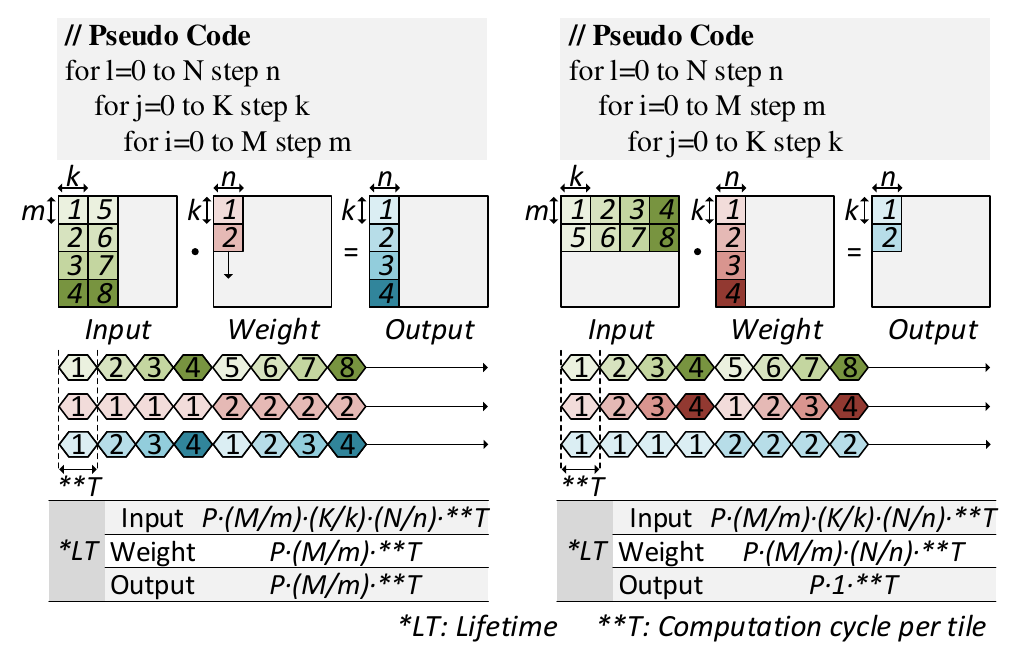}
\caption{Lifetime of Data per Tiling Scheme Analysis}
\vspace{-0.1in}
\label{figure9}
\end{figure}

\textbf{Step 2: Analyze lifetime of data.} In a tiled GEMM operation, the lifetime of each tile varies according to the tiling scheme used. Figure ~\ref{figure9} illustrates this change through two pseudo-code examples and their corresponding computation flows for tiled GEMM. While the overall computation time remains consistent, the lifetime of each tile differs, as the equations show. The computation cycle per tile, denoted by $T$, changes depending on the processing type (i.e., bit-serial and bit-parallel). In step 2, the lifetime of each data is analyzed across all tiling schemes identified in step 1.

\textbf{Step 3: Estimate energy consumption.} In step 3, we perform energy modeling by using the lifetime analyzed in step 2 and the memory specifications of our proposed reconfigurable eDRAM (i.e., RBL voltage swing, power consumption, and retention time table) as input. For each tiling scheme, total energy consumption is estimated ($E_{total}$) for various RBL voltage swings as Equation (1). $E_{PIM}$ and $E_{Buffer}$ denote the energy consumption of the PIM macro operation and buffer access, respectively. The $E_{PIM}$ and $E_{Buffer}$ are calculated as in Equation (2) and (3). $E_{Acc}$, $E_{PU}$, and $E_{Ref}$ represent the energy consumption of eDRAM access, processing unit, and refresh per operation, respectively. $N$ denotes the number of operations. $T_{Life}$ and $T_{Retention}$ refer to the lifetime of mapped data and the retention time of eDRAM. The prefixes $P$ and $B$ identify the PIM macro and buffer.

\begin{scriptsize}
\begin{align}
    E_{total} = E_{PIM}+E_{Buffer}
\end{align}
\begin{align}
    E_{PIM} = (E_{P\_Acc}+E_{PU}) \times P\_N + E_{P\_Ref} \times \left[\frac{T_{P\_Life}}{T_{P\_Retention}}\right]
\end{align}
\begin{align}
    E_{Buffer} = (E_{B\_Acc}) \times B\_N + E_{B\_Ref} \times \left[\frac{T_{B\_Life}}{T_{B\_Retention}}\right]
\end{align}
\end{scriptsize}

\textbf{Step 4: Find optimal computation flow and memory operation.} Based on the output of the energy modeling, tiling scheme, and memory operation that minimize overall energy consumption are identified. By forwarding the results to the PIM macro controller, we fully optimize energy consumption of the memory.

\begin{figure*}[t]
\centering
\includegraphics[width=1\textwidth]{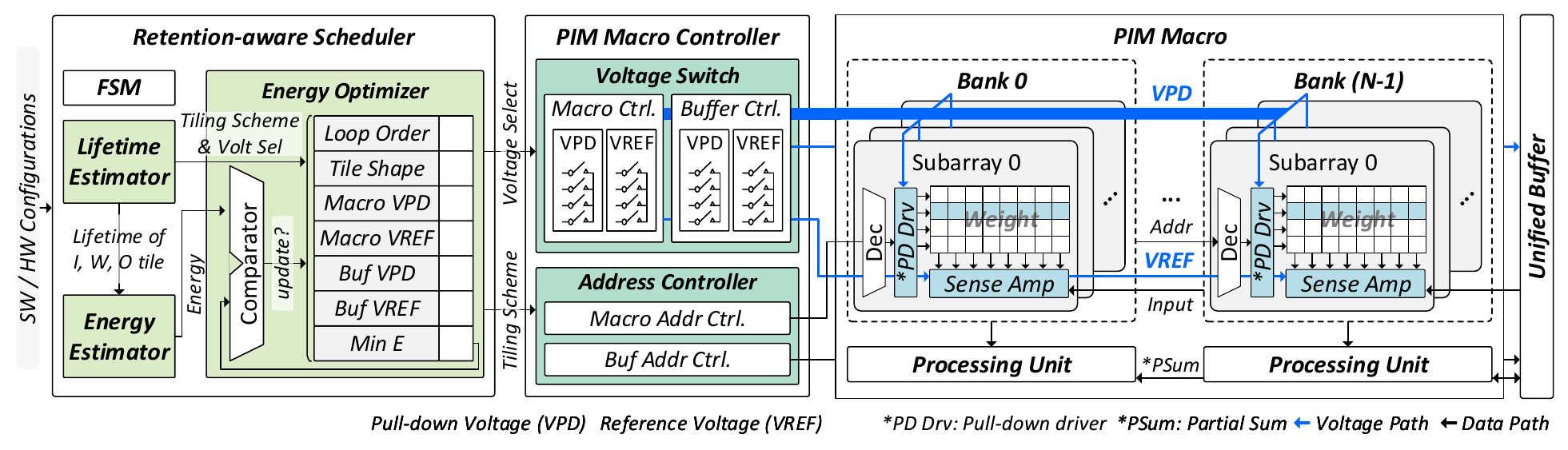}
\caption{\sysname's Hardware Template Architecture}
\label{figure6}
\end{figure*}

\subsection{Hardware Template Architecture}

As highlighted in Section ~\ref{observations}, a memory design that does not take into account the property of an architecture and its use cases falls short of achieving high energy efficiency. To address this problem, we propose a reconfigurable eDRAM-based PIM hardware template architecture to achieve adaptability to the outputs of the \sysname's scheduling phase. Figure ~\ref{figure6} illustrates \sysname’s hardware template architecture, which consists of a retention-aware scheduler, a PIM macro controller, a PIM macro, and a unified buffer. The user can change the memory configuration (e.g., the subarray shape, the number of subarrays and banks, and the buffer size) and the processing unit configuration (e.g., operation types such as dot product/row-wise, processing type, and bit precision).

The retention-aware scheduler finds the optimal operating conditions by evaluating the energy consumption across all possible tiling schemes and memory operations for the given architecture configuration, as detailed in Section ~\ref{retention-aware scheduling}. The lifetime estimator calculates the lifetime of the input, weight, and output tile for the chosen tiling scheme. Subsequently, the energy estimator computes the energy consumption for each memory operation. The energy optimizer forwards the most energy-efficient tiling scheme and voltage select signal to the PIM macro controller. The controller modulates the memory access patterns and adjusts eDRAM operations by activating the row that stores the weight and broadcasting the input to the subarray’s sense amplifier. The input is then multiplied by the weight in the sense amplifier and passed to the processing unit. This unit executes the GEMM or GEMV operation using the output from the eDRAM subarray. Throughout this process, the PIM macro controller adjusts the eDRAM operation by tuning the pull-down voltage (VPD) and reference voltage (VREF), marked with the blue line. The following section delves into the specific hardware implementation for the reconfigurable eDRAM. A further optimization scheme to minimize redundant refresh in eDRAM is discussed below.

Reducing memory access power inevitably leads to more frequent refresh, which in turn increases energy consumption and latency overhead. To counter this, we employ the refresh skipping scheme for the data that has a shorter lifetime than the retention time or the data that is no longer utilized in computation. For example, it is possible to skip refresh if the lifetime of the partial sum is shorter than the retention time, as new data replaces the old. Input tiles and weight tiles also no longer need to be refreshed when they are no longer used in the computation. By skipping redundant refresh in these cases, the PIM macro controller effectively mitigates the increased refresh overhead, thereby enhancing overall energy efficiency.

\subsection{Reconfigurable eDRAM}
\begin{figure}[t]
\centering
\includegraphics[width=0.48\textwidth]{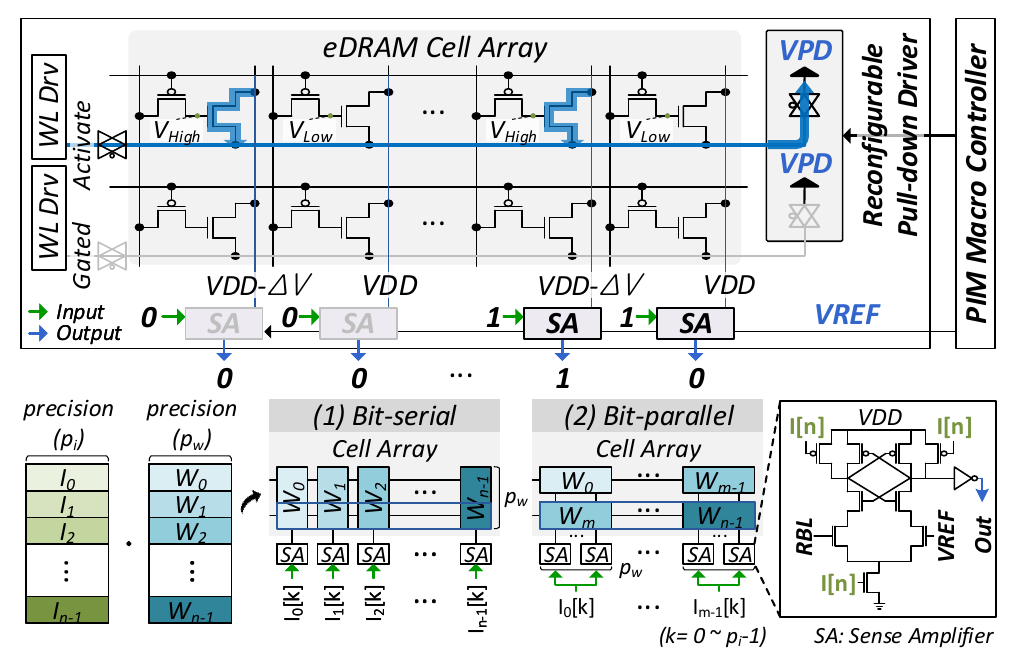}
\caption{Reconfigurable 2T eDRAM Circuit Design}
\vspace{-0.1in}
\label{figure7}
\end{figure}

We propose a reconfigurable 2T eDRAM macro capable of adjusting the RBL voltage swing according to the pull-down voltage (VPD) controlled by the PIM macro controller. Additionally, we implement two optimizations-adjusting the reference voltage and employing sense amplifier power gating-to further enhance energy efficiency. Figure ~\ref{figure7} illustrates the detailed hardware implementation of the macro, featuring a reconfigurable pull-down driver. This driver employs a pass transistor, targeting a VPD dictated by the PIM macro controller instead of VSS. Adjusting the VPD closer to VSS increases the RBL voltage swing, which in turn enhances a sensing margin and extends a retention time. However, this adjustment concurrently increases the power consumption of the memory access. 

\begin{figure*}[t]
\centering
\includegraphics[width=1\textwidth]{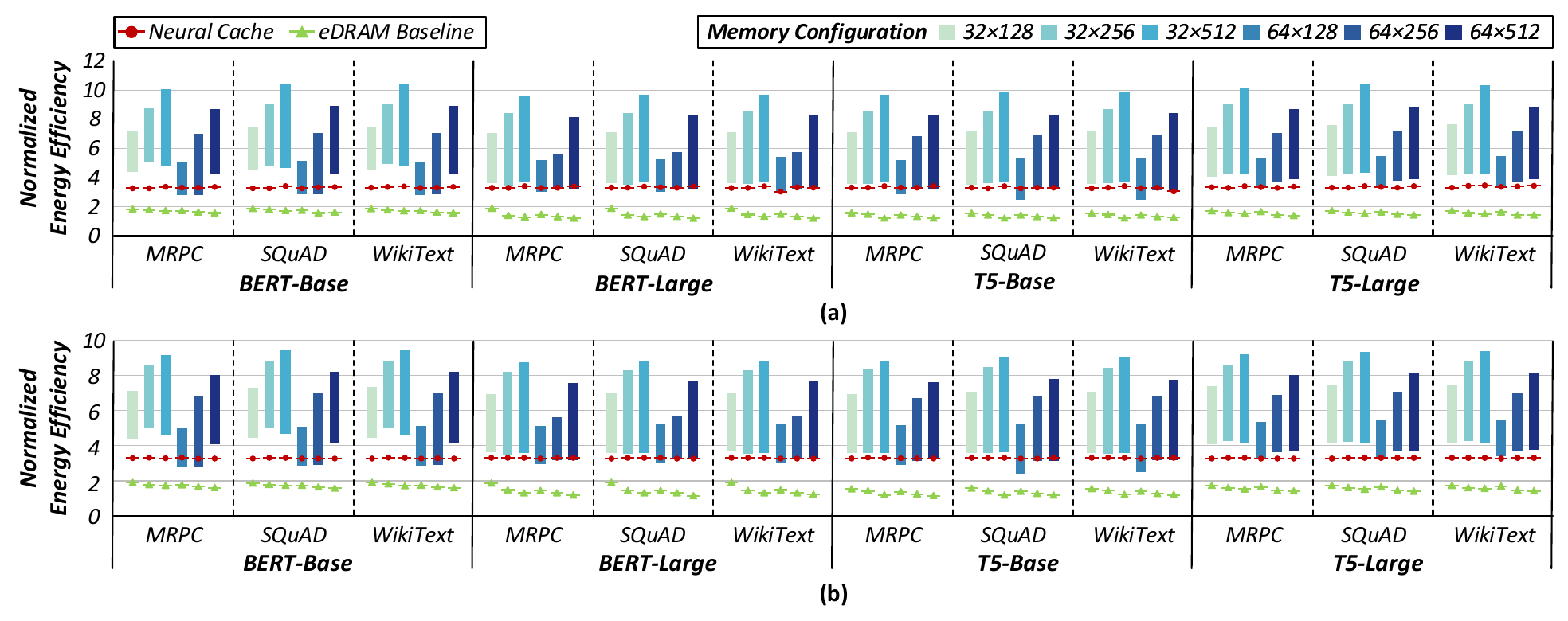}
\caption{The \sysname Framework’s Normalized Energy Efficiency Comparison with eDRAM Baseline and Neural Cache for (a) Bit-serial Processing Unit and (b) Bit-parallel Processing Unit}
\label{figure_energy_efficiency_1}
\end{figure*}

Before runtime, the \sysname framework estimates overall energy consumption for various VPD settings. This estimation is enabled by forwarding the specification about the power consumption and the retention time relative to the VPD of our proposed eDRAM macro to the energy modeling of the scheduling phase. The \sysname framework is able to identify optimal tiling scheme and memory operation to achieve the highest energy efficiency by pre-estimating the energy consumption at each VPD level. The following describes additional optimization schemes aimed at augmenting the functionality and power efficiency of our proposed eDRAM design.

The functionality of the sense amplifier and the retention time of the memory are significantly affected by the reference voltage, as the sense amplifier utilizes this voltage to convert the RBL voltage swing into a digital signal. Thus, we opt to use the midpoint of the RBL voltage swing between data 1 and data 0 as the reference voltage to optimize the performance of the sense amplifier and the retention time. This reference voltage is pre-calculated for each level of VPD, enabling the PIM macro controller to select the reference voltage corresponding to the VPD.

We employ the sense amplifier power gating to mitigate the trade-off between the RBL voltage swing and the power consumption of the sense amplifier, which also effectively substitutes the function of a 1-bit multiplier. As the VPD is set higher, the power consumption of the sense amplifier increases due to a reduced sensing margin. With our design, if the input for computation with weight is zero, the sense amplifier is power gated, ensuring it outputs data 0 regardless of the data actually stored, as depicted in Figure ~\ref{figure7}. Depending on how the data is mapped to the eDRAM cell array, the input data is broadcast in different ways. For bit-parallel mapping, a shared input bit is broadcast because each weight data bit operates with the same input bit. Conversely, in bit-serial mapping, each weight bit operates with a different input bit requiring individual broadcasting. The output of the sense amplifier is the same as the output of the 1-bit multiplier, so it can be aggregated directly in the processing unit. Thus, sense amplifier power gating accomplishes two benefits: 1) it reduces both the area and power overhead of the 1-bit multiplier, and 2) it further optimizes memory access.

\section{Evaluation}
\label{evaluation}

\subsection{Experimental Setup}
\subsubsection{Benchmark}
To evaluate the energy efficiency of our \sysname framework, we use four Transformer models: BERT-Base, BERT-Large, T5-Base, and T5-Large. We utilize MRPC, SQuAD2.0, and WikiText-103 for datasets, each with an input token length of 1024. The data types of activation and weight data are configured to INT8. To evaluate our proposed sense amplifier power gating, we use bit-wise sparsity of activation for each layer.

\begin{figure}[t]
\centering
\includegraphics[width=0.48\textwidth]{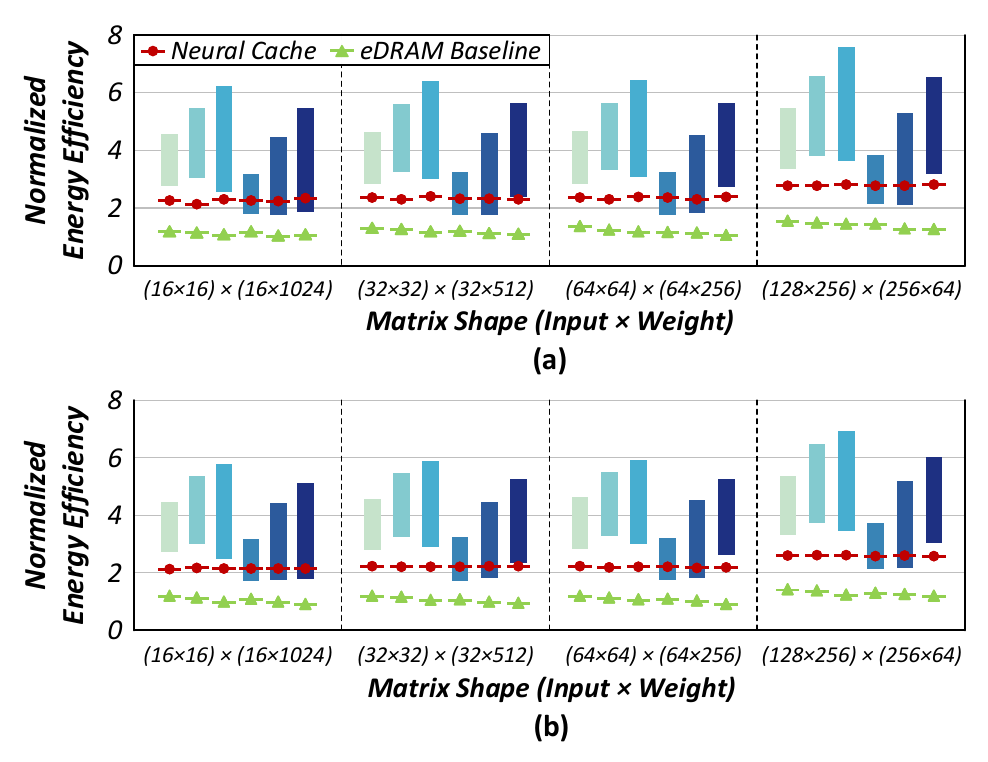}
\caption{The \sysname Framework’s Normalized Energy Efficiency with Different Matrix Configurations for (a) Bit-serial Processing Unit and (b) Bit-parallel Processing Unit}
\vspace{-0.1in}
\label{figure_energy_efficiency_2}
\end{figure}

\subsubsection{Hardware Configuration}
To validate reconfigurable eDRAM, we measure the power consumption across six eDRAM subarray configurations: 32$\times$128, 32$\times$256, 32$\times$512, 64$\times$128, 64$\times$256, and 64$\times$512. We evaluate our framework with a 60KB unified buffer and 16KB PIM macro. The PIM macro features four banks, each configured with a number of subarrays to achieve a 4KB capacity per bank. We use a weight stationary data flow in our architecture, processing inputs in a bit-serial manner. For Weight, we evaluate two methods: bit-serial and bit-parallel. The processing unit consists of an adder tree and an accumulator for MAC operation.

\subsubsection{Simulation Methodology}

We develop C++ based in-house simulator for the scheduling phase of \sysname, incorporating the power consumption of retention-aware scheduler, PIM macro controller, eDRAM access, and processing unit. It estimates overall energy consumption and energy efficiency. The energy modeling of the hardware template are from post-layout simulation in HSPICE for eDRAM memory, and from Synopsys Design Compiler for scheduler, controller, and processing unit with Samsung’s 28nm technology. The hardware operates under conditions of a 200MHz frequency, 1V voltage, and room temperature. We simulate our reconfigurable eDRAM across four different VPDs: 200mV, 300mV, 400mV, and 500mV.

Since this is the first framework that optimizes the energy consumption of eDRAM macro as a digital PIM device to the best of our knowledge, we do not compare it against previous analog circuit-based eDRAM PIMs. Instead, we compare the \sysname framework with two baselines: an eDRAM baseline and Neural Cache ~\cite{eckert2018neural}. Firstly, we use an eDRAM baseline operating under fixed memory operation without any of the proposed optimization schemes. We configure the baseline to maintain the highest RBL voltage swing, aiming to minimize the refresh overhead as normal memory design usually adopts. Secondly, we compare our framework against Neural Cache, prior SRAM-based PIM, because previous digital PIM macros usually adopt SRAM as memory. We scale Neural Cache to the same operational conditions for consistency.

\begin{figure}[t]
\centering
\includegraphics[width=0.48\textwidth]{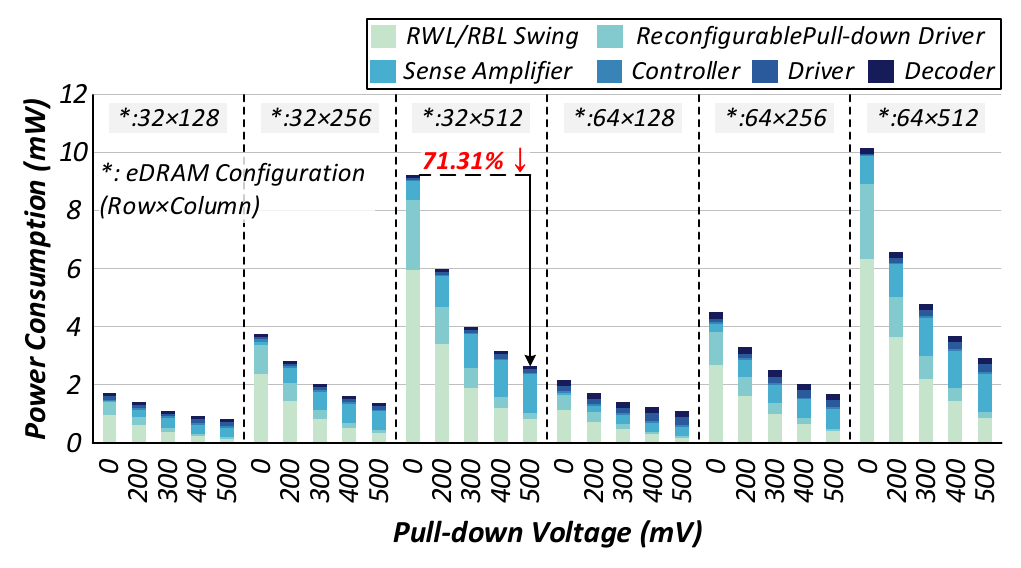}
\caption{Power Consumption Breakdown and Reduction}
\vspace{-0.1in}
\label{figure_breakdown}
\end{figure}

\subsection{Energy Efficiency of \sysname}
\textbf{Comparison with various transformer models}
Figure ~\ref{figure_energy_efficiency_1} shows the normalized energy efficiency with four Transformer models and three benchmarks for six memory configurations, using a processing unit with (a) bit-serial operators and (b) bit-parallel operators. Energy efficiency values are normalized with the minimum value of the eDRAM baseline. The graph shows the range from maximum to minimum normalized energy efficiency of all possible operating conditions configured by the \sysname framework. Notably, the framework consistently outperforms both the eDRAM baseline and Neural Cache in the best case identified through retention-aware scheduling. These performance improvements come from \sysname’s capability to identify the optimal scheme through retention-aware scheduling and control memory operation optimally for the actual use case. Furthermore, mitigating the increased overhead from the sense amplifier and refresh fully optimizes the energy consumption of eDRAM macro, ensuring peak energy efficiency. There are some cases where the worst-case energy efficiency falls below that of Neural Cache despite the lower access power of eDRAM compared to SRAM. This is because suboptimal memory control and tiling schemes lead to increased refresh overhead, underscoring the importance of memory control and scheduling in adopting eDRAM as a PIM device. With the bit-serial processing unit, the \sysname framework achieves 1.59$\times$-2.66$\times$, 1.52$\times$-3.05$\times$, and 2.87$\times$-8.16$\times$ higher energy efficiency than worst point, Neural Cache, and eDRAM baseline, respectively, at the optimal point. The bit-parallel processing unit demonstrates performance enhancements similar to those of the bit-serial processing unit, as both processing units with the same throughput exhibit similar operating power.

\begin{figure}[t]
\centering
\includegraphics[width=0.48\textwidth]{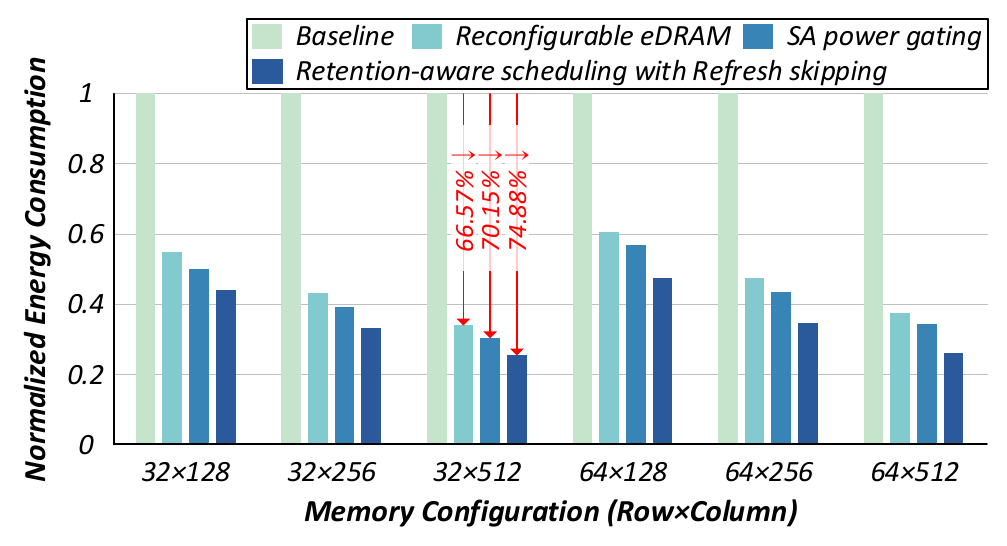}
\caption{Energy Reduction per Scheme}
\vspace{-0.1in}
\label{figure_optimization}
\end{figure}

\textbf{Comparison with different matrix configurations}
Figure ~\ref{figure_energy_efficiency_2} illustrates the impact of varying input and weight matrix shapes on the energy efficiency across six memory configurations with the BERT-Base and MRPC dataset. While changing the matrix shape influences the lifetime of the tile and thereby affects tiling schemes, our framework adeptly identifies the most energy-efficient operating conditions. Regardless of the workload, \sysname consistently finds the optimal tiling scheme and memory operation by utilizing the SW configuration as input and conducting energy modeling before runtime. With the bit-serial processing unit, our framework achieves energy efficiency improvements ranging from 1.63$\times$ to 2.93$\times$, 1.52$\times$ to 2.95$\times$, and 2.87$\times$ to 5.85$\times$ compared to the worst point, Neural Cache, and eDRAM baseline, respectively. Similarly, with the bit-parallel processing unit, the framework achieves 1.62$\times$-2.78$\times$, 1.51$\times$-2.76$\times$, and 2.84$\times$-5.38$\times$ higher energy efficiency than worst point, Neural Cache, and eDRAM baseline, respectively.

\subsection{Power Consumption Comparison}

Figure ~\ref{figure_breakdown} shows the power consumption of memory access and its breakdown for the five different VPDs with the six memory configurations, respectively. The data shows that higher VPD reduces power consumption of the RWL/RBL voltage swing and the pull-down driver, resulting in a lower power consumption. However, a higher VPD level results in increased power consumption by the sense amplifier. For example, the power consumption of memory access reduces by 71.31\% when VPD elevates from 0 to 500mV for eDRAM macro with a cell array size of 32$\times$512. However, this adjustment causes the sense amplifier’s portion in total power consumption to increase from 6.82\% to 50.14\%. This shift indicates the need to optimize the sense amplifier’s power consumption to reduce memory access power further.

\subsection{Energy Consumption Optimization}
Figure ~\ref{figure_optimization} shows changes of the eDRAM macro energy consumption normalized to the eDRAM baseline by gradually adding each of the proposed schemes. Each indicates reconfigurable eDRAM, sense amplifier power gating, and retention-aware scheduling with refresh skipping, respectively. As shown in the graph, since memory access energy consumption is dominant in PIM, we can optimize energy consumption by 54.04\% on average by reducing memory access power with reconfigurable eDRAM. Furthermore, addressing the increased power consumption of the sense amplifier with power gating reduces the energy consumption by 58.00\% on average.

Although the reconfigurable eDRAM can reduce memory access power by up to 71.31\%, the energy optimization only by the eDRAM is 66.57\%. This is because the refresh overhead is far more increased due to the shortened retention time. The scheduling phase of the \sysname framework enables the best use of our proposed hardware design by identifying the optimal tiling scheme and memory operation through energy modeling. Employing this retention-aware scheduling, the \sysname framework effectively optimizes the energy consumption of eDRAM macro compared to the baseline by 56.43\%, 66.95\%, 74.88\%, 52.97\%, 65.76\%, and 74.28\% in each memory configurations.

\begin{figure}[t]
\centering
\includegraphics[width=0.48\textwidth]{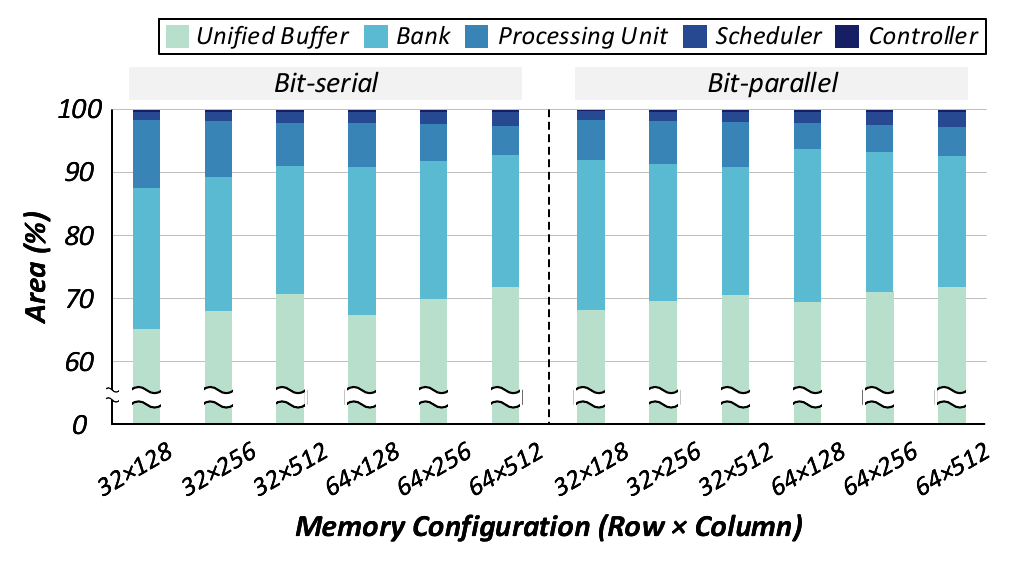}
\caption{Area Breakdown of the \sysname Framework}
\label{figure_area_breakdown}
\end{figure}

\subsection{Area and Energy Breakdown}
The \sysname framework's area breakdown is depicted in Figure ~\ref{figure_area_breakdown}. The graph shows the breakdown when using bit-serial and bit-parallel operators as processing units for each of the six memory configurations. As the graph shows, for both types of processing units, the proposed retention-aware scheduler and PIM macro controller take up less than 2.5\% and 1\% of the total area, respectively, that is marginal. We can see that most of the area overhead comes from the unified buffer (around 70\%) and bank (around 20\%). The proposed reconfigurable driver takes up 13.7\%, 8.41\%, 4.74\%, 17.24\%, 10.58\%, and 5.97\% of the subarray area for the six configurations, respectively.

\begin{figure}[t]
\centering
\includegraphics[width=0.48\textwidth]{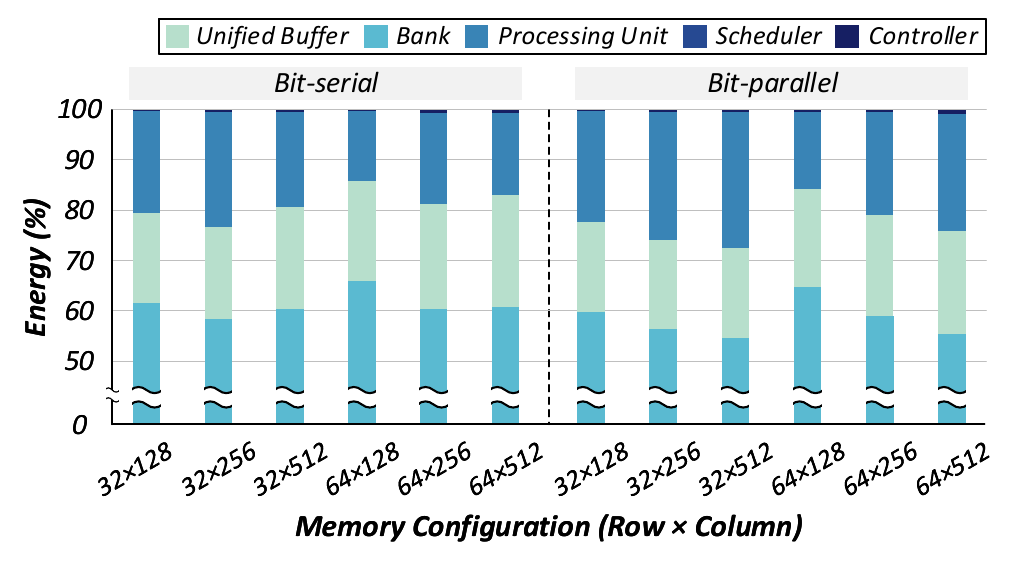}
\caption{Energy Breakdown of the \sysname Framework}
\vspace{-0.1in}
\label{figure_energy_breakdown}
\end{figure}

Figure ~\ref{figure_energy_breakdown} shows the energy breakdown for the same experimental conditions. The data shows that the retention-aware scheduler and PIM macro controller both account for less than 1\% energy consumption in the overall operation, regardless of the type of processing unit. The PIM macro, which performs the actual computation, accounts for most of the energy consumption, with the bank accounting for about 60\% and the processing unit accounting for about 20\%.

\section{Discussion}
\label{discussion}

\textbf{Appliance to different models and dataset}
Even though we only evaluate the proposed \sysname framework for Transformer models and datasets in this study, the framework is equally applicable to various memory-bound applications such as deep neural networks (DNN), recommendation systems, and other workloads that include GEMM or GEMV. Unlike previous PIM designs that use memory as a static device, our framework can achieve the highest energy efficiency with the proposed reconfigurable pull-down driver to dynamically control the memory operation, considering the properties of the workload assigned to the PIM macro. In other words, by using the workload configurations to be assigned to the PIM macro and the configurations of the memory and processing unit as input for retention-aware scheduling, our framework can find the optimal operating conditions to achieve maximum energy efficiency.

\textbf{Scalability of the \sysname framework}
In our evaluation, we explore various memory configurations by adjusting the number of subarrays per bank while maintaining a total memory capacity. Notably, our framework is also able to accommodate any total memory capacity and configuration of subarrays and banks because our approach includes a data mapping to the PIM macro, ensuring that operations can be processed in parallel and subarrays are controlled equally. Such mapping is a common consideration in PIM macro design, ensuring that memories and processing units share the same control signals and input data within the macro.


\textbf{Estimation with different processing unit}
Our experiments assess two types of processing units: bit-serial and bit-parallel operators. The \sysname framework can be extended to other operators by incorporating the operating power of the operators as input in the energy modeling during the scheduling phase. However, for operators that process both input and weight in a bit-parallel manner, their placement in each subarray presents challenges due to its high area and power overhead. Instead, these types of operators are typically configured so that they are shared across multiple subarrays, such as at the bank level. This arrangement allows data from various subarrays to be multiplexed and directed to the shared operator. By integrating these practical constraints into our energy modeling and hardware configuration, we can identify optimal operating conditions in the same manner.

\section{Conclusion}
\label{conclusion}

In this work, we propose \sysname, an energy optimization framework for eDRAM-based PIM. \sysname employs retention-aware scheduling that is able to identify the most energy-efficient tiling scheme and memory operation. We design a novel reconfigurable eDRAM-based architecture that can be dynamically controlled according to the outcomes of the retention-aware scheduling. As a result, \sysname framework is able to reduce memory access power by up to 71.31\% and achieve energy efficiency improvements up to 3.05$\times$ over Neural Cache. With its minimal area and energy overhead for the scheduler and controller, \sysname offers a solution for developing energy-efficient eDRAM-based PIM applications.

\bibliographystyle{unsrt}
 \bibliography{Reference.bib}

\end{document}